\documentclass{article}

    \usepackage[final, nonatbib]{neurips_2024}

\usepackage[utf8]{inputenc} 
\usepackage[T1]{fontenc}    
\usepackage{hyperref}       
\usepackage{url}            
\usepackage{booktabs}       
\usepackage{amsfonts}       
\usepackage{nicefrac}       
\usepackage{microtype}      
\usepackage{xcolor}         

\usepackage{times}
\usepackage{epsfig}
\usepackage{graphicx}
\usepackage{amsmath}
\usepackage{amssymb}

\usepackage{subfigure}

\usepackage{caption}
\usepackage{wrapfig}
\usepackage{multirow}
\usepackage{makecell}
\usepackage{pifont}
\usepackage[normalem]{ulem}

\newcommand{\eg}{\textit{e.g.}}
\newcommand{\ie}{\textit{i.e.}}

\newcommand{\wrt}{\textit{w.r.t.~}}

\title{Improved Generation of Adversarial Examples Against Safety-aligned LLMs}

\author{
Qizhang Li$^{1,2}$, Yiwen Guo$^3$\thanks{Yiwen Guo leads the project and serves as the corresponding author.}\ \,, Wangmeng Zuo$^1$, Hao Chen$^4$  \\
\small{$^1$Harbin Institute of Technology,\, $^2$Tencent Security Big Data Lab,\, $^3$Independent Researcher,\, $^4$UC Davis}\\
\small{\texttt{\{liqizhang95,guoyiwen89\}@gmail.com\quad wmzuo@hit.edu.cn\quad chen@ucdavis.edu}}
}

\begin{document}

\maketitle

\begin{abstract}
Adversarial prompts (or say, adversarial examples) generated using gradient-based methods exhibit outstanding performance in performing automatic jailbreak attacks against safety-aligned LLMs.
Nevertheless, due to the discrete nature of texts, the input gradient of LLMs struggles to precisely reflect the magnitude of loss change that results from token replacements in the prompt, leading to limited attack success rates against safety-aligned LLMs, even in the \emph{white-box} setting.
In this paper, we explore a new perspective on this problem, suggesting that it can be alleviated by leveraging innovations inspired in transfer-based attacks that were originally proposed for attacking \emph{black-box} image classification models.
For the first time, we appropriate the ideologies of effective methods among these transfer-based attacks, \ie, Skip Gradient Method~\cite{Wu2020} and Intermediate Level Attack~\cite{Huang2019}, into gradient-based adversarial prompt generation and achieve significant performance gains without introducing obvious computational cost.
Meanwhile, by discussing mechanisms behind the gains, new insights are drawn, and proper combinations of these methods are also developed.
Our empirical results show that $87\%$ of the query-specific adversarial suffixes generated by the developed combination can induce Llama-2-7B-Chat to produce the output that exactly matches the target string on AdvBench. This match rate is $33\%$ higher than that of a very strong baseline known as GCG, demonstrating advanced discrete optimization for adversarial prompt generation against LLMs.
In addition, without introducing obvious cost, the combination achieves $>30\%$ absolute increase in attack success rates compared with GCG when generating both query-specific ($38\% \rightarrow 68\%$) and universal adversarial prompts ($26.68\% \rightarrow 60.32\%$) for attacking the Llama-2-7B-Chat model on AdvBench.
Code at: \href{https://github.com/qizhangli/Gradient-based-Jailbreak-Attacks}{https://github.com/qizhangli/Gradient-based-Jailbreak-Attacks}.
\end{abstract}

\section{Introduction}
\label{sec:intro}

Large language models (LLMs) have demonstrated a formidable capacity for language comprehension and the generation of human-like text.
Safty-aligned LLMs, refined through specific fine-tuning mechanisms~\cite{ouyang2022training, bai2022constitutional, korbak2023pretraining, glaese2022improving}, are anticipated to yield responses that are not only helpful but also devoid of harm in response to user instructions.
However, certain studies~\cite{shen2023anything, wei2024jailbroken, zou2023universal, perez2022red, chao2023jailbreaking, liu2023autodan} reveal that these models have not yet achieved perfect alignment. 
It has been demonstrated that these models can be carefully prompted to produce harmful content through the introduction of meticulously crafted prompts, a phenomenon known as ``jailbreak''~\cite{wei2024jailbroken}.
The manually designed jailbreak prompts are crafted by carefully constructing scenarios that mislead the models, necessitating a significant amount of work.
In contrast, adversarial examples are automatically generated with the intent deceiving models to generate harmful responses, presenting a more insidious challenge to model robustness. 

One of the main difficulties in generating adversarial examples for NLP models lies in the fact that text is discrete by nature, making it challenging to use gradient-based optimization methods to devise adversarial attacks.
There has been some work~\cite{guo2021gbda, wen2024pez, shin2020autoprompt, jones2023automatically, zou2023universal} attempted to overcome this issue.
For instance, recently, a method called Greedy Coordinate Gradient (GCG) attack~\cite{zou2023universal} has shown significant jailbreaking improvements, by calculating gradients of cross-entropy loss \wrt one-hot representations of chosen tokens in a prompt and replacing them in a greedy manner.
However, due to the fact that the gradients \wrt one-hot vectors do not provide precise indication of the loss change that results from a token replacement, the GCG attack shows limited white-box attack success rates against some safty-aligned LLMs, \eg, Llama-2-Chat models~\cite{touvron2023llama}.

In this paper, we carefully examine the discrepancy between the gradient of the adversarial loss \wrt one-hot vectors and the real effect of the change in loss that results from token replacement. 
We present a new perspective that this gap resembles the gap between input gradients calculated using a substitute model and the real effect of perturbing inputs on the prediction of a black-box victim model, which has been widely studied in transfer-based attacks against black-box image classification models~\cite{Szegedy2014, Papernot2016transferability, Liu2017, Huang2019, Wu2020, guo2020back, li2024theoretical, li2023towards}.
Based on this new perspective, for the first time, we attempt to appropriating the ideologies of two effective methods among these transfer-based methods, \ie, Skip Gradient Method (SGM)~\cite{Wu2020} and Intermediate Level Attack (ILA)~\cite{Huang2019}, to improve the gradient-based attacks against LLMs. 
With appropriate adaptations, we successfully inject these ideologies into the gradient-based adversarial prompt generation without additional computational cost.
By discussing the mechanisms behind the advanced performance, we provide new insights about improving discrete optimizations on LLMs.
Moreover, we provide an appropriate combination of these methods.
The experimental results demonstrate that $87\%$ of the query-specific adversarial suffixes generated by the combination for attacking Llama-2-7B-Chat on AdvBench can induce the model output exact target string, which outperforms a strong baseline named GCG attack (54\%), indicating an advanced discrete optimization for adversarial prompt generation against LLMs.
In addition, the combination achieves attack success rates of 68\% for query-specific and 60.32\% for universal adversarial prompt generation when attacking Llama-2-7B-Chat on AdvBench, which are higher than those of the baseline GCG (38\% and 26.68\%).

\section{Related Work}
\label{sec:related_work}

\textbf{Jailbreak attacks}. 
Recent work highlights that the safety-aligned models are still not perfectly aligned~\cite{carlini2024aligned, shen2023anything, wei2024jailbroken}, the safety-aligned LLMs can be induced to produce harmful content by some carefully designed prompts, known as jailbreak attacks~\cite{wei2024jailbroken}. 
This has raised security concerns and attracted great attention.
In addition to some manually designed prompt methods~\cite{wei2024jailbroken, shen2023anything}, numerous automatic jailbreak attack methods have been proposed. Some methods directly optimize the text input through gradient-based optimization~\cite{wallace2019universal, guo2021gbda, shin2020autoprompt, wen2024pez, jones2023automatically, zou2023universal}. Another line of work involves using LLMs as optimizers to jailbreak the victim LLM~\cite{perez2022red, chao2023jailbreaking, mehrotra2023tree}. There are also methods that focus on designing special jailbreaking templates or pipelines~\cite{liu2023autodan, shah2023scalable, casper2023explore, deng2023jailbreaker, zeng2024johnny, wei2023jailbreak}.
According to the knowledge of the victim model, these methods can also be divided into white-box attacks and black-box attacks. 
In the context of white-box attacks, the attackers have full access to the architecture and parameters of the victim LLM, making them can leverage the gradient with respect to the inputs. 
As demonstrated by recent benchmark~\cite{mazeika2024harmbench}, represented by a current method as known as GCG attack~\cite{zou2023universal}, gradient-based automatic jailbreak attacks have shown the most powerful performance in compromising LLMs in the setting of white-box attack. 
However, due to the discrete nature of text input, dealing with the discrete optimization problem is rather challenging, which limits the success rates of attacks, especially those against Llama-2-Chat models~\cite{touvron2023llama}.
In our work, we primarily focus on solving the discrete optimization problem in gradient-based automatic jailbreak attacks to improve the success rate in the white-box setting.

\textbf{Transfer-based attacks}.
Transfer-based attacks attempt to craft adversarial examples on a white-box substitute model to attack the black-box victim model, by leveraging the transferability of adversarial examples~\cite{Szegedy2014}, which is a phenomenon that adversarial examples crafted on a white-box substitute model can also mislead the unknown victim models with a decent success rate.
The transfer-based attacks have been thoroughly investigated in the setting of attacking image classification models~\cite{Kurakin2017, Xie2019, Dong2019, Wu2020, Huang2019, gubri2022lgv, guo2020back, guo2022intermediate, li2023making, li2023improving, li2024improving, li2023towards}.
Some recent methods also utilize the transferability of adversarial prompts to perform black-box attacks against LLMs~\cite{zou2023universal, liu2023autodan, sitawarin2024pal}.
While in this work, we mainly focus on improving the white-box attack success rate.
In our work, we reveal a closely relationship between the optimization in transfer-based attacks and discrete optimization of the gradient-based jailbreak attacks against LLMs. We then appropriate the ideologies of two effective transfer-based attack methods developed in the setting of attacking image classification model, \ie, SGM~\cite{Wu2020} and ILA~\cite{Huang2019}. 
Moreover, by adapting these strategies and analyzing the mechanism behind them, we provide some new insights about potential solutions for addressing problems involving discrete optimization in NLP models with transformer architecture, \eg, prompt tuning.

\section{Gap Between Input Gradients and the Effects of Token Replacements}
\label{sec:3}
In this section, we first discuss the main obstacle in achieving effective gradient-based attacks on safety-aligned LLMs, which is considered as the gap between input gradients and the effects of token replacements, and then we show how transfer-based strategies can be adapted to overcome the issue.

\subsection{Rethinking Gradient-based Attacks Against LLMs}
\label{sec:3.1}
Previous endeavors utilize the gradient \wrt the token embeddings~\cite{lester2021power, wen2024pez} or \wrt the one-hot representations of tokens~\cite{ebrahimi2017hotflip, shin2020autoprompt, zou2023universal, jones2023automatically}, in order to solve the discrete optimization problem efficiently during attacking LLMs. 
A recent method, named Greedy Coordinate Gradient (GCG)~\cite{zou2023universal}, shows a significant improvement over other optimizers (\eg, AutoPrompt~\cite{shin2020autoprompt} and PEZ~\cite{wen2024pez}) on performing gradient-based attacks against LLMs in the white-box setting. 
Due to its effectiveness, we take GCG as a strong baseline and as a representative example to analyze previous gradient-based attacks against LLMs in this section.

A typical LLM $f: \mathcal{X} \rightarrow \mathbb{R}^{|V|}$ with a vocabulary $V$ is trained to map a sequence of tokens $x_{1:n} = [x_1, ..., x_n] \in \mathcal{X}, x_i \in V$ to a probability distribution over the next token, denoted as $p_f(x_{n+1} | x_{1:n})$.
To evoke a jailbreak to induce the a safety-aligned LLM generate harmful content according the user query, GCG attempts to add an adversarial suffix to the original user query and iteratively modifies the adversarial suffix to encourages the model output an affirmative target phrase, \eg, ``Sure, here's ...''.
Consider an adversarial prompt (which is the concatenation of a user query and an adversarial suffix) as $x_{1:n}$ and a further concatenation with a target phrase as $x_{1:n^*}$, GCG aims to minimize the adversarial loss $L(x_{1:n^*})$ (denoted as $L(x)$ for simplicity), which corresponds to the negative log probability of the target phrase. It can be written as 
\begin{equation}\label{eq:gcg_obj}
\begin{aligned}
    \min_{x_{\mathcal{A}}\in\{1, \dots, V\}^{|\mathcal{A}|}} L(x_{1:n^*}) = \min_{x_{\mathcal{A}}\in\{1, \dots, V\}^{|\mathcal{A}|}} \frac{1}{n^*-n} \sum_{i=1}^{n^*-n} - \log p_f(x_{n+i} | x_{1:n+i-1}),
\end{aligned}
\end{equation}
where $x_{\mathcal{A}}$ denotes the tokens of adversarial suffix in $x_{1:n}$ and $\mathcal{A}$ denotes the set of indices of the adversarial suffix tokens.
At each iteration, it computes the gradient of the adversarial loss \wrt the one-hot vector of a single token and uses each value of gradient to estimate the effect of replacing the current token with the corresponding one on loss.
Since the discrete nature of input, there is a gap between input gradients and the effects of token replacements, thus the estimate is not accurate enough.
Previous efforts attempt to solve this problem by collecting a candidate set of token replacements according to the Top-$k$ values in the gradient and evaluating the candidates to pick the best token replacement with minimal loss~\cite{zou2023universal, shin2020autoprompt}.
Due to the large gap, they requires a large candidate set size, \eg, 512 in GCG, which result in a large computational cost.
Ideally, if the input gradients can accurately reflect the effects of token replacements, $k=1$ is sufficient to achieve minimal loss without needing a candidate set, thus obtaining the optimal token replacement with the lowest computational cost.
Therefore, we advocate for refining the input gradient to narrow the gap, thereby improving performance while reducing computational cost.

We attempt to introduce a new perspective to the gap between the input gradients and the real effects of token replacements.
Let us first revisit transfer-based attacks on image classification models. 
To generate an adversarial example that misleads an unknown victim model, where the input gradient is not accessible, attackers use a white-box substitute model as a proxy for the victim model and utilize its input gradient.
Due to the gap between the input gradient of the substitute model and the real input gradient of the victim model, directly using the input gradient of the substitute model to modify the example yields unsatisfactory results. 
Efforts~\cite{Dong2018Boosting, Xie2019, Huang2019, Wu2020, li2020yet, guo2020back, li2023making} have been made to refine the gradient computed on the substitute model in order to narrow this gap.
Returning to our discrete optimization problem in the adversarial prompt generation, the strategy of utilizing the gradient \wrt the one-hot vector of the token is actually treating the one-hot vector as though it was a continuous variable, resulting in the gap between this gradient and the ``real gradient'' (\eg, the real effects of token replacements). 
We consider that this gap is analogous to the input gradient gap between substitute and victim models in the context of transfer-based attacks.
This new perspective allows one to introduce a series of innovations developed within the realm of transfer-based attacks on image classification models to refine the gradient computation in the context of gradient-based adversarial prompt generation against safety-aligned LLMs.

From the results of a recent benchmark of these transfer-based attacks~\cite{li2023towards}, we can observe that several strategies, including SGM~\cite{Wu2020}, SE~\cite{naseer2021improving}, PNA~\cite{wei2022towards}, and a series of intermediate level attacks (ILA~\cite{Huang2019}, ILA++~\cite{guo2022intermediate}, FIA~\cite{wang2021feature}, and NAA~\cite{zhang2022improving}), are obviously effective when generating adversarial examples on models with a transformer architecture.
Among them, PNA ignores the backpropagation of the attention map and SE requires considerable ability to directly predict the probability from the intermediate representations, thus their strategies are difficult to be adapted to the context of LLM attacks. 
Therefore, we consider drawing inspiration from SGM and ILA (which is the representative of all intermediate level attacks).

\subsection{Reducing the Gradients from Residual Modules}
\label{sec:3.2}

\begin{figure}[t]
    \begin{minipage}{0.2\textwidth}
    \centering
    \vskip0.08in
    \hspace{2em}\includegraphics[width=1\textwidth]{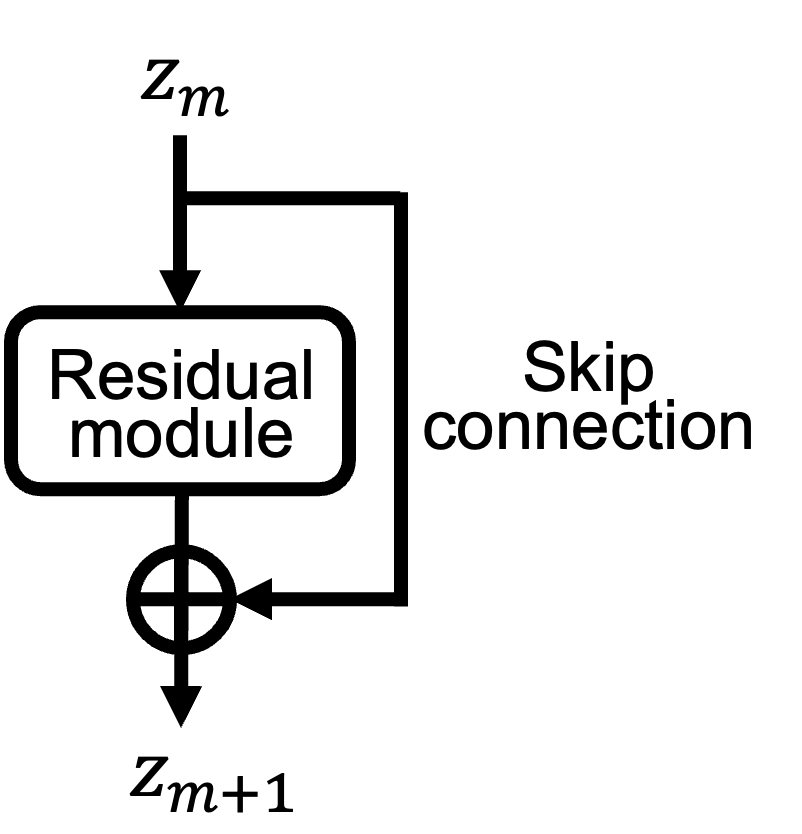}\vskip0.16in
    \caption{An example of the residual block. }
    \label{fig:block}
    \end{minipage}
    \hfill
    \begin{minipage}{0.78\textwidth}
    \centering
    \vskip-0.15in
    \subfigure[Loss]{
     \includegraphics[width=0.49\textwidth]{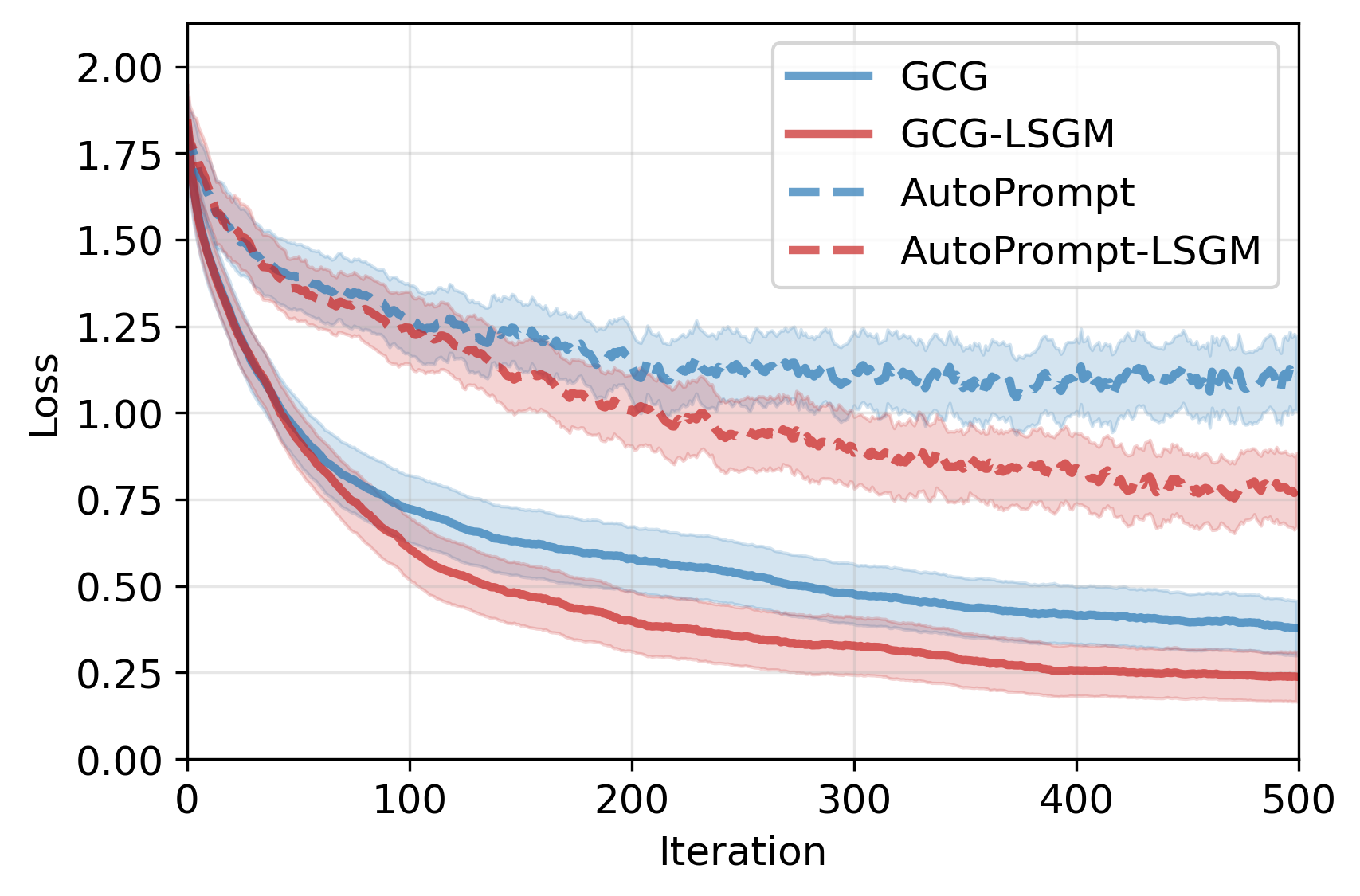}}
     \subfigure[Match rate]{
     \includegraphics[width=0.475\textwidth]{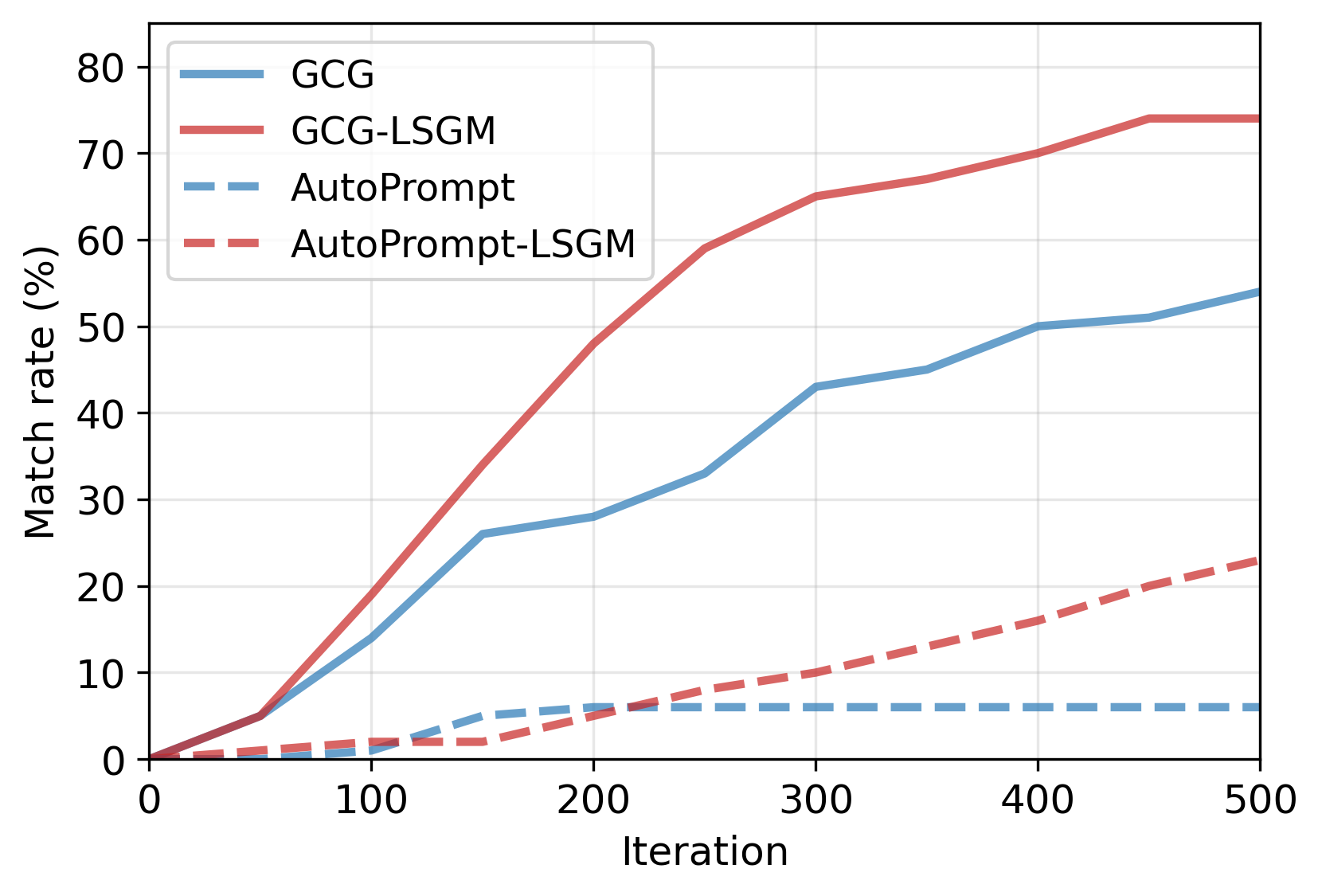}} 
    \vskip-0.1in
    \caption{How (a) the loss and (b) the match rate changes with attack iterations. The attacks are performed against Llama-2-7B-Chat model to generate query-specific adversarial suffixes on AdvBench. Best viewed in color.}
    \label{fig:sgm}
    \end{minipage}
    \vskip-0.2in
\end{figure}


Modern deep neural networks typically comprise a number of residual blocks, each consisting of a residual module and a skip connection branch, as depicted in Figure~\ref{fig:block}.
SGM~\cite{Wu2020} experimentally found that reducing the gradients from residual modules during backpropagation can improve the transfer-based attacks against image classification models, indicating that it can reduce the gap between the input gradients and the effects resulting from perturbing inputs on a unknown victim model.
In this section, we investigate whether the strategy of reducing gradients from residual modules can also enhance gradient-based attacks against LLMs in a white-box setting. Additionally, we discuss the mechanisms behind this strategy to provide new insights.

An $l$-layers LLM can be decomposed into $2l$ residual building blocks, with each block consisting of a residual module (which should be an MLP or an attention module) and a parallel skip connection branch as illustrated in Figure~\ref{fig:block}. 
The $m$-th block maps an intermediate representation $z_m$ to $z_{m+1}$, \ie, $z_{m+1} = I(z_{m}) + R_{m} (z_{m})$, where $I$ is an identity function representing the skip connection branch and $R_{m}$ denotes the residual module of the $m$-th block.
By adopting a decay factor $\gamma \in [0,1]$ for the residual modules, SGM calculates the derivative of the $m$-th block as $\nabla_{z_m} z_{m+1} = 1 + \gamma \nabla_{z_{m}} R_m(z_m)$.
We incorporate this strategy into gradient-based automatic adversarial prompt generation, denoted as Language SGM (LSGM).
We evaluate the performance of integrating this strategy into GCG and AutoPrompt attacks by setting $\gamma=0.5$, and show the results in Figure~\ref{fig:sgm}. 
The experiment is conducted to generate query-specific adversarial suffixes against Llama-2-7B-Chat~\cite{touvron2023llama} on the first 100 harmful queries in AdvBench~\cite{zou2023universal}. 
To provide a more comprehensive comparison, we report not only the adversarial loss but also the fraction of adversarial prompts that result in outputs exactly matching the target string, dubbed match rate, which is also used as an evaluation metric in the paper of GCG~\cite{zou2023universal}. 
It can be seen from the figure that reducing gradients from residual modules can indeed improve the performance of gradient-based adversarial prompt generation.
The GCG-LSGM achieves a match rate of $72\%$, while the baseline (\ie, GCG) only obtains $54\%$ match rate.
We also evaluate the attack success rate (ASR) by using the evaluator proposed by HarmBench~\cite{mazeika2024harmbench}. It shows an ASR of $62\%$ when using GCG-LSGM, while the GCG only achieves $38\%$. 
On the other hand, when reducing the gradient from the skip connection, both the match rate and the attack success rate will drop to 0\%.
The results confirm that reducing gradients from residual modules helps the gradients \wrt one-hot representations to be more effective to indicate the real effects of token replacements.

\begin{figure}[t]
    \begin{minipage}{0.45\textwidth}
    \centering
    \vskip-0.2in
    \hspace{-1em}\includegraphics[width=0.76\columnwidth]{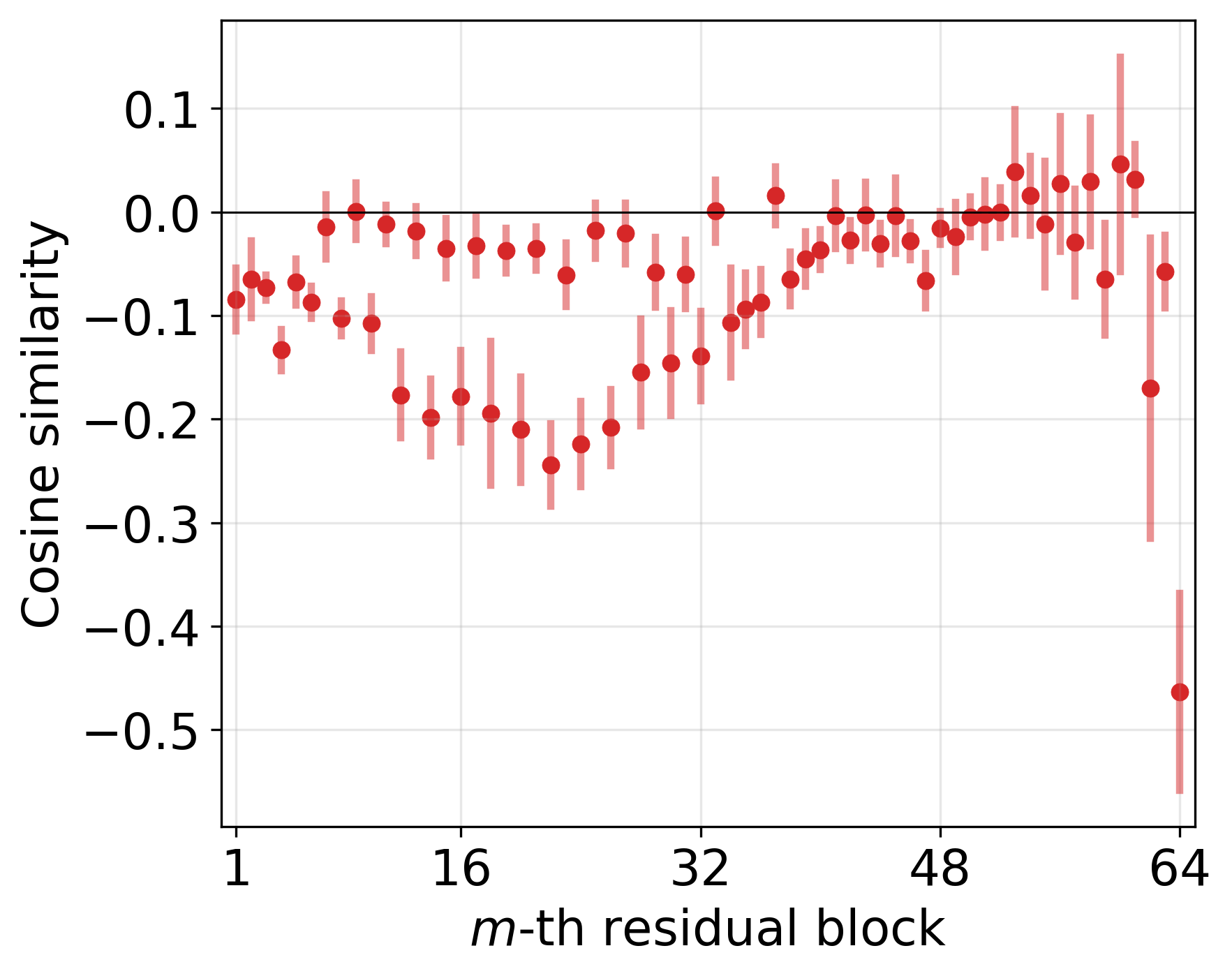}
    \caption{The cosine similarities between the gradients from residual modules and the gradients from skip connections in different residual blocks.}
    \label{fig:cos}
    \end{minipage}
    \hfill
    \begin{minipage}{0.52\textwidth}
    \centering
    \vskip-0.2in
    \includegraphics[width=\columnwidth]{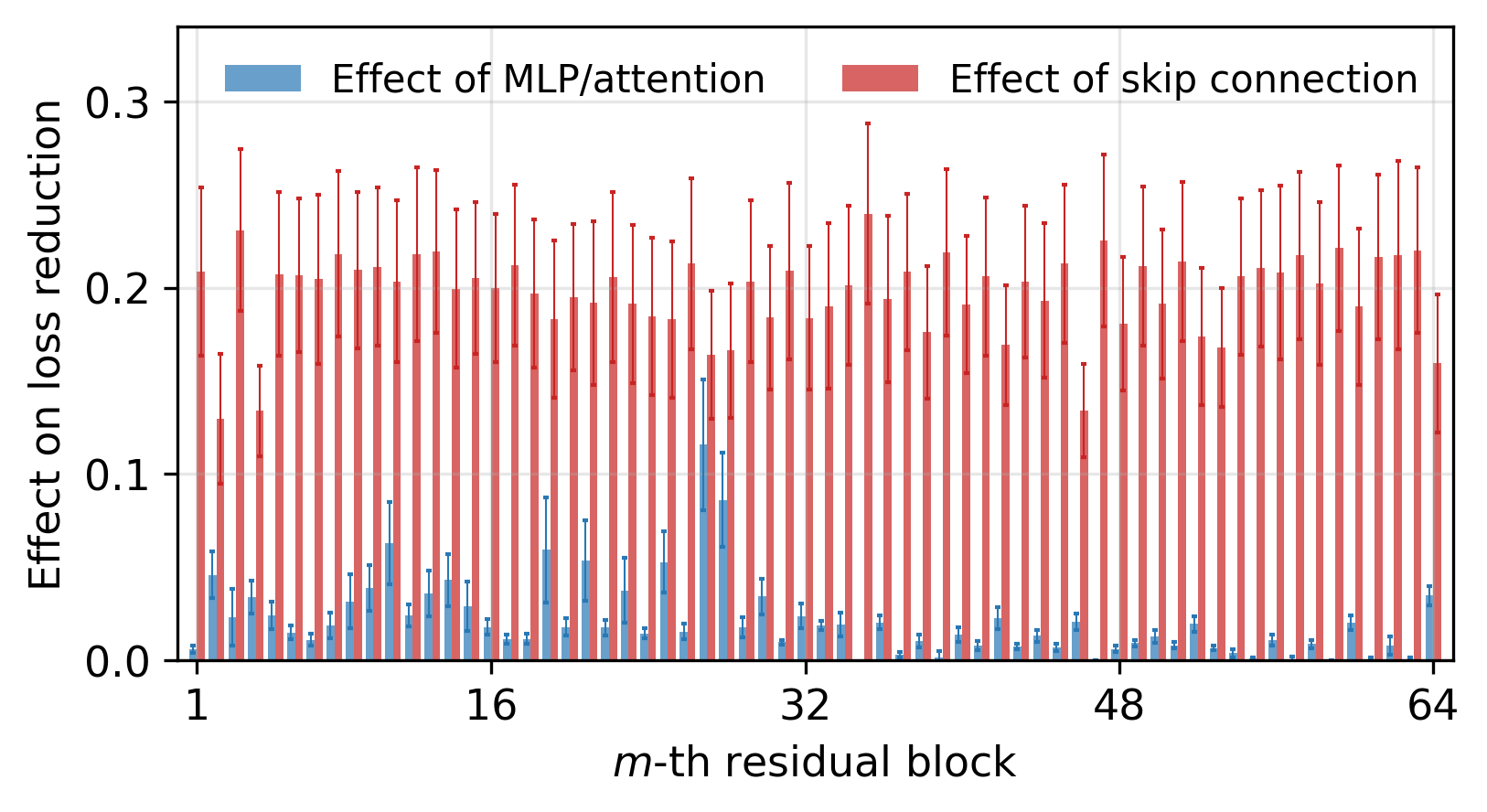}
    \caption{Comparing the average effects of residual modules and the average effects of skip connections on the change in adversarial loss varies with different residual blocks. Best viewed in color.}
    \label{fig:effect}
    \end{minipage}
    \vskip-0.1in
\end{figure}

Having seen the effectiveness of GCG-LSGM, let us further delve deep into the mechanism behind the method.
We begin by analyzing the computational graph and gradient flow of a building block.
Following the chain rule, the gradient of adversarial loss \wrt $z_m$ can be written as a summation of two terms: $\nabla_{z_m} L(x) = \nabla_{z_{m+1}} L(x) + \nabla_{z_{m}} R(z_m) \nabla_{z_{m+1}} L(x)$. 
The first term represents the gradient from the skip connection, and the second term represents the gradient from the residual module.
They represent the impact on the loss caused by changes in $z_{m}$ through skip connection branch and residual module branch, respectively.
In Figure~\ref{fig:cos}, we visualize the average cosine similarity between these two terms at the 100-th iteration of the GCG attack against Llama2-7B-Chat across 100 examples. 
The same observations can be obtained at other iterations during the GCG attack.
Somewhat surprisingly, it can be seen that these two terms have negative cosine similarity in most blocks during the iterative optimization. 
Obviously, the summation of two directionally conflicting gradients may mitigate the effect of each other's branch on reducing the loss.  
Hence, reducing the gradients from residual modules achieves lower loss values by sacrificing in the effects of the residual modules to trade more effects of the skip connection on loss.
The success of GCG-LSGM somehow implies the gradient flowing from skip connections better reflect the effect of token replacements on adversarial loss.

To confirm the conjecture, we attempt to evaluate the effect of each branch towards reducing the loss. 
Inspired by the causal tracing technique~\cite{meng2022locating}, we perform the following steps to compute the effect of each branch's hidden state.
First, we give an adversarial prompt to the model to obtain the loss, denoted by $L(x)$.
Second, we randomly alter a token from the adversarial prompt and record the hidden states in the two branches, \ie, $I(\Tilde{z}_{m})$ and $R_{m} (\Tilde{z}_{m})$, given the altered adversarial prompt.
Finally, we feed the original adversarial prompt into the model and modify the forward computation by replacing $I(z_m)$ with $I(\Tilde{z}_m)$ (or $R_{m} (z_{m})$ with $R_{m} (\Tilde{z}_{m})$ ), to obtain the modified loss $\Tilde{L}(x)$.
The effect of a branch is represented by the difference between the loss of obtained on the third step and the first step, \ie, $\Tilde{L}_m(x) - L(x)$. 
A high $\tilde{L}_m(x) - L(x)$ indicates the branch to replaced hidden state is important on affect the adversarial loss.
By averaging the values $\Tilde{L}_{m}(x) - L(x)$ over a collection of adversarial prompts, we can get the average effects of residual module and skip connection in the $m$-th block.
In Figure~\ref{fig:effect}, we show the average effects of residual modules and skip connections. The adversarial promp are obtained by performing GCG attack against Llama-2-7B-Chat model on AdvBench.
It can be seen that the skip connections show much greater effects than residual modules on the adversarial loss, which confirms the conjecture.
The observation further implies that the effects of adversarial information primarily propagate through the skip connection branch within each residual block. 
This observation, alongside the superior gradient-based attack performance of LSGM, further suggests that certain discrete optimization challenges within LLMs, \eg, prompt tuning, might be more effectively addressed by understanding the information flow throughout the forward pass.

\subsection{Adapting Intermediate Level Attack for Gradient-base Attacks Against LLMs}
\label{sec:3.3}

Intermediate level attacks~\cite{Huang2019, guo2019simple, wang2021feature,zhang2022improving,li2023improving} opt to maximizing the scalar projection of the intermediate level representation onto a ``directional guide'' for generating non-target adversarial examples to attack image classification models.
As a representative method, ILA~\cite{Huang2019} defines the directional guide as the intermediate level representation discrepancy between the adversarial example obtained by a preliminary attack, \eg, I-FGSM~\cite{Kurakin2017}, and corresponding benign example.
In this section, we first examine whether the strategy of ILA can be directly applied to the gradient-based attacks on LLMs. Then, by seeing the failure of direct application, we investigate the proper way to adapt this strategy to the gradient-based attacks on LLMs.

Let $h_r$ be the intermediate representation at the $r$-th layer, and the directional guide $v_r=\hat{h}^0_r - \hat{h}^t_r$ is obtained by a $t$-iterations preliminary attack (\eg, GCG).
Note that since our objective is to minimize the adversarial loss instead of maximizing the victim's classification loss, as in the original setting of ILA paper~\cite{Huang2019}, the definition of the directional guide is the opposite of that in the ILA paper.
According the implementation of ILA, we maximize the scalar projection of the intermediate representation $h_r$ onto the directional guide $v_r$, \ie, maximize $L_{\mathrm{ILA}} (x) = h^T_{r} v_r$.
However, we evaluated ILA using GCG as the back-end method and found that it exhibited deteriorated performance compared to the baseline.
Specifically, when attacking Llama-2-7B-Chat, it achieves a match rate of $44\%$, while the baseline GCG obtains $54\%$ match rate.
This result demonstrates that when applied directly, the ILA fails to facilitate gradient-based adversarial prompt generation.

\begin{figure}[t]
\centering 
 \subfigure[PCCs computed on the $h_r$ ]{\label{fig:5.1}
 \hspace{-0.5em}\includegraphics[width=0.36\textwidth]{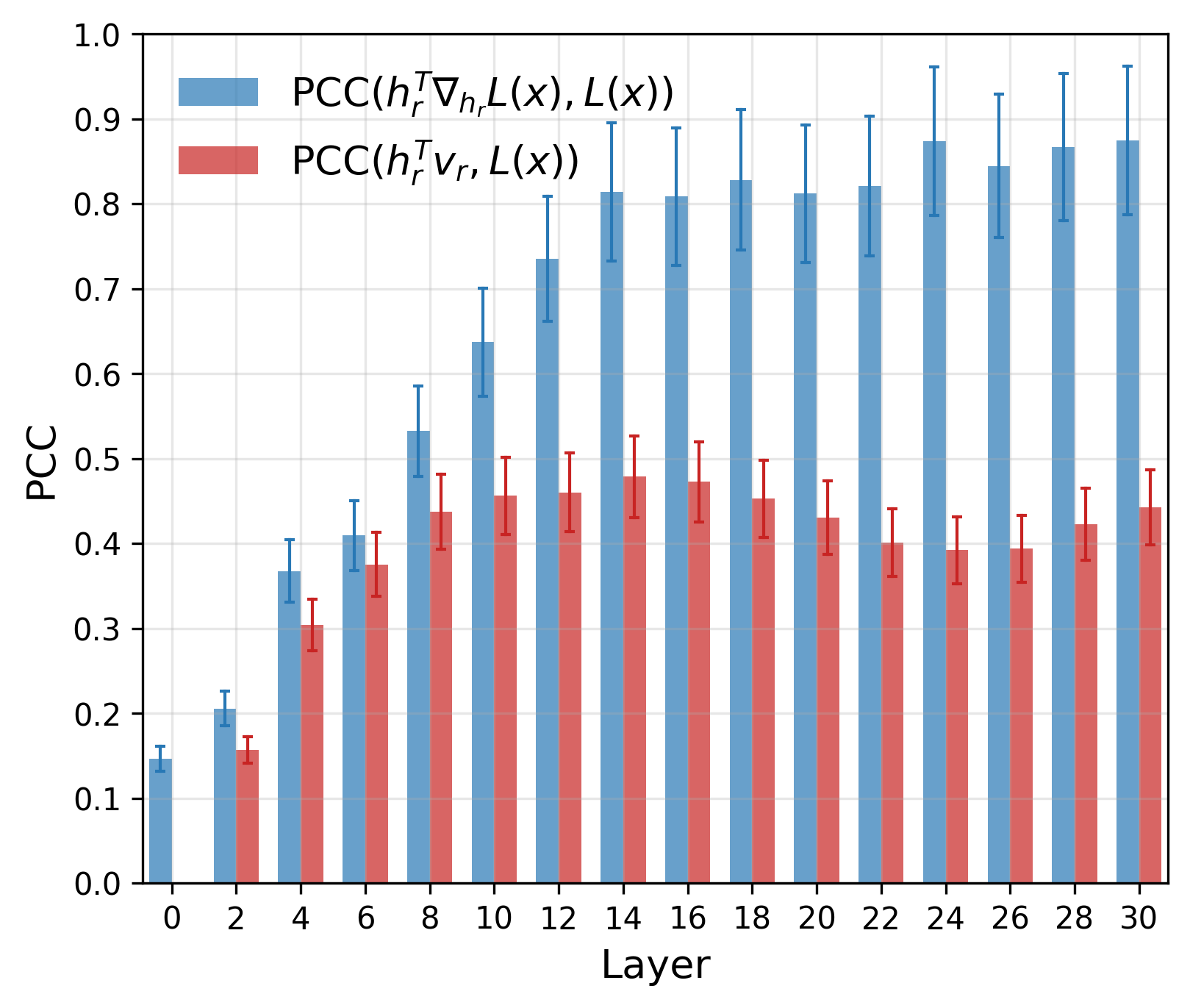}}\hspace{-0.5em}
 \subfigure[PCCs computed on the $h_{r,o}$]{\label{fig:5.2}
 \includegraphics[width=0.63\textwidth]{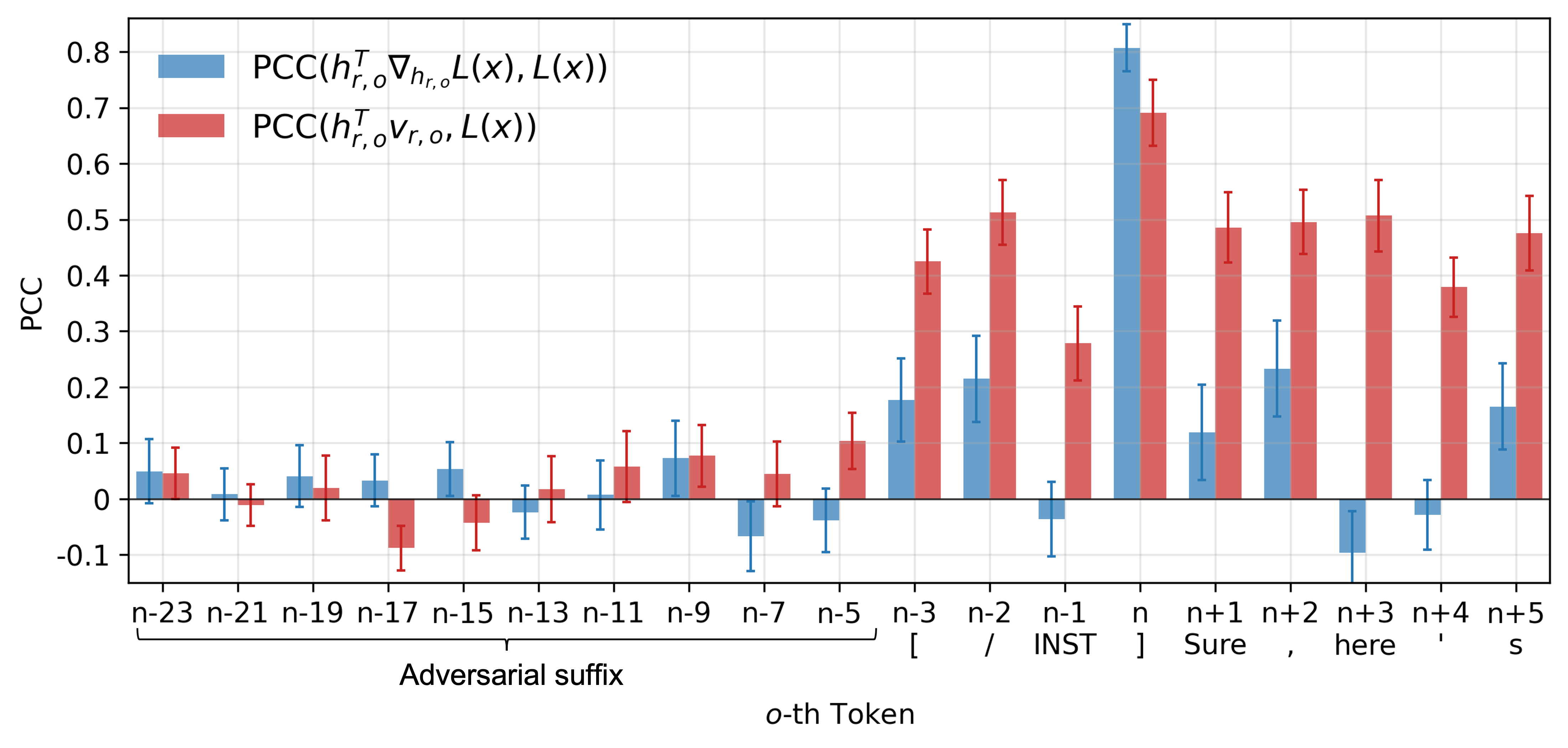}} 
\caption{(a) The PCCs computed on the entire intermediate representations, \ie, $\mathrm{PCC} (h^T_r \nabla_{h_r}L(x), L(x))$ and $\mathrm{PCC} (h^T_r v_r, L(x))$, at different layers of Llama-2-7B-Chat.
(b) The PCCs computed on the $o$-th token intermediate representations, \ie, $\mathrm{PCC} (h_{r,o}\nabla_{h_{r,o}}L(x), L(x))$ and $\mathrm{PCC} (h^T_{r,o} v_{r,o}, L(x))$, at the mid-layer of Llama-2-7B-Chat. Best viewed in color.}\vskip-0.15in
\end{figure}

Let us discuss why directly introducing ILA fails to improve performance.
Recall that intermediate-level attacks against image classification models essentially assume that the scalar projection of an intermediate representation onto the directional guide has a positive correlation with the transferability of adversarial examples, which means that the larger the scalar projection obtained, the greater the transferability (\ie, higher classification loss on victim models) the adversarial example achieves~\cite{li2023improving}.
Previous work~\cite{li2023improving} has also shown that the ILA is equivalent to replacing the gradient \wrt the intermediate representation with the directional guide. 
This demonstrates that ILA necessitates a stronger positive correlation between the scalar projection of the representation onto the directional guide and the victim's classification loss, compared to the positive correlation between the scalar projection of the representation onto its gradient and the victim's classification loss.
Back to our setting, we conduct experiments to examine whether these assumptions hold when attacking LLMs.
We use Pearson's correlation coefficient (PCC) to show the correlation between the scalar projection and adversarial loss.
With a range in $[-1,1]$, a PCC close to $1$ indicates a positive correlation, while a PCC close to $-1$ means a negative correlation.
The experiment is conducted with following steps.
Firstly, we perform a GCG attack to obtain an adversarial example, and use it to derive a directional guide $v_r$ and the gradient of adversarial loss $L(x)$ \wrt the intermediate representation $h_r$, \ie,\ $\nabla_{h_r}L(x)$.
Next, we randomly alter adversarial tokens several times and input these new prompts into the model to collect a set of intermediate representations paired with their corresponding adversarial losses.
We then calculate the PCC between the scalar projection of the intermediate representation onto the directional guide and the adversarial loss, \ie, $\mathrm{PCC} (h^T_r v_r, L(x))$, and the PCC between the scalar projection of the intermediate representation onto its gradient and the adversarial loss, \ie, $\mathrm{PCC} (h^T_r \nabla_{h_r}L(x), L(x))$.
We perform this experiment using Llama-2-7B-Chat on AdvBench.
In Figure~\ref{fig:5.1}, we show the PCCs between the scalar projection and the adversarial loss at different layers.
It can be seen that in every layer, the PCCs computed using the directional guide are all $\leq 0.5$, showing a weaker correlation compared to the PCCs computed using the intermediate gradient, which exceed 0.8 after the 14th layer.
This confirms that the failure to directly apply ILA is due to its poorer performance in the intermediate layers in indicating the change of adversarial loss compared to the baseline.
This phenomenon is distinct from the one observed in the setting of attacking image classification models~\cite{li2023improving}.

\begin{figure}[t]
    \begin{minipage}{0.28\textwidth}
    \centering
    \includegraphics[width=1\textwidth]{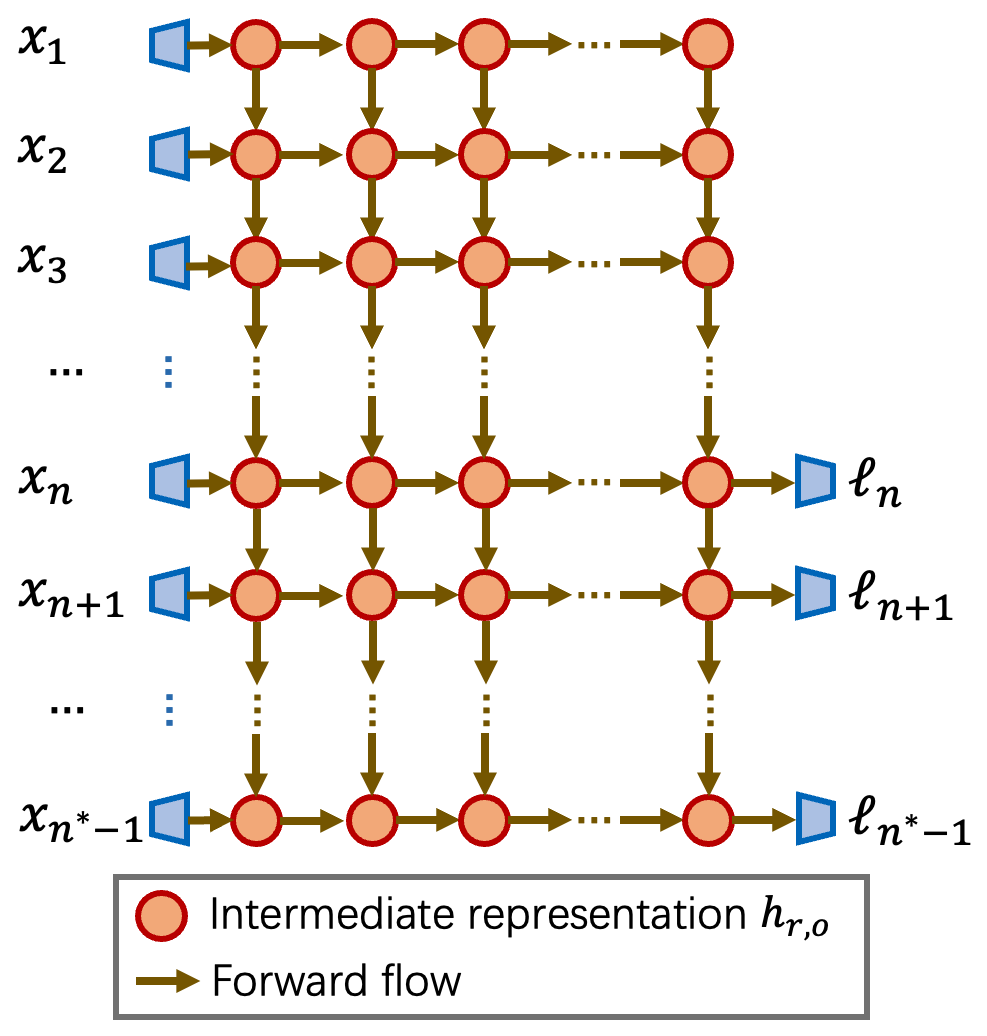}
    \caption{An example of the internal computation of an LLM. The $\ell$ represents cross-entropy loss.}
    \label{fig:grid}
    \end{minipage}
    \hfill
    \begin{minipage}{0.7\textwidth}
    \centering
    \subfigure[Loss]{
     \includegraphics[width=0.49\textwidth]{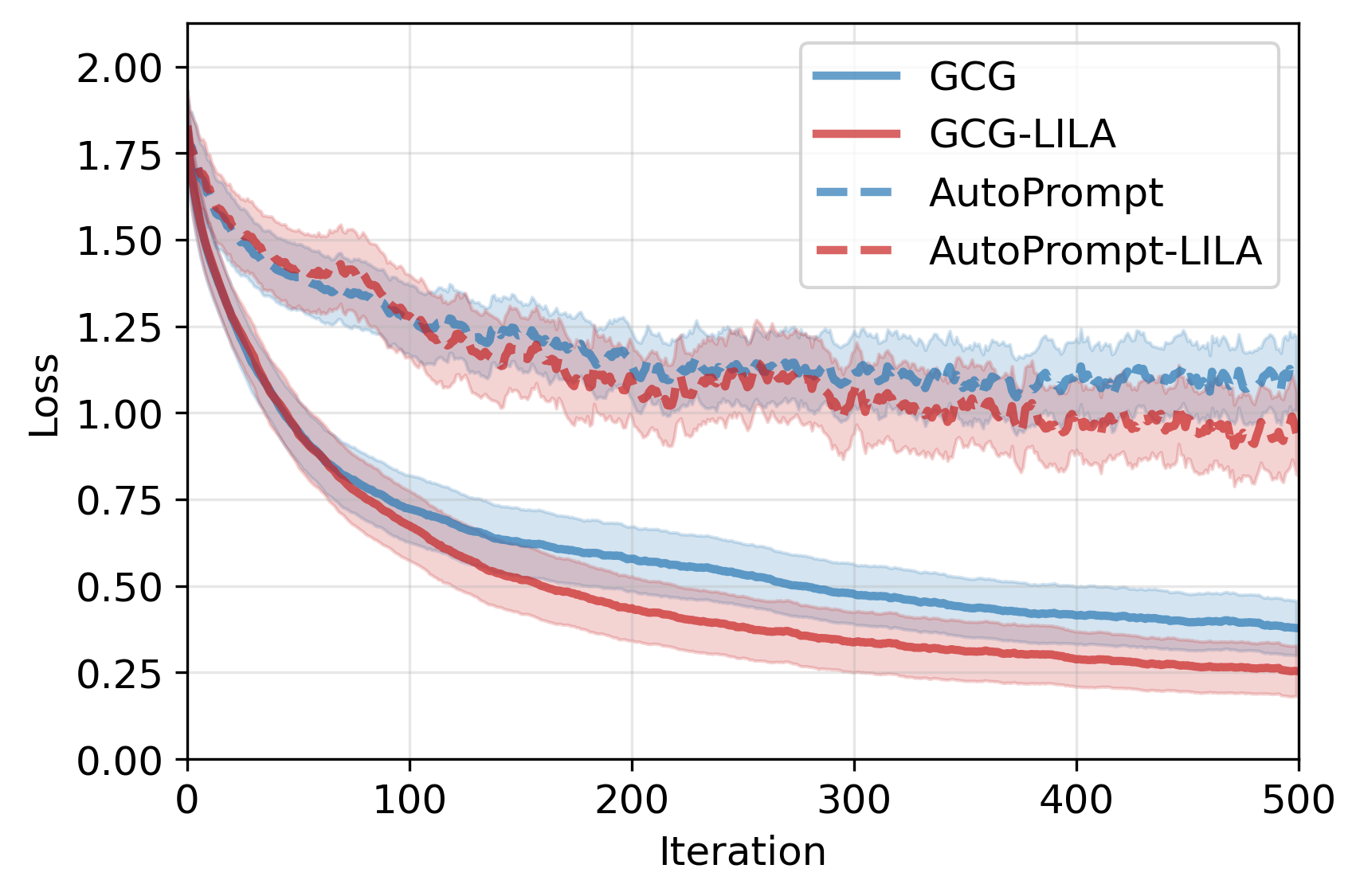}}
     \subfigure[Match rate]{
     \includegraphics[width=0.475\textwidth]{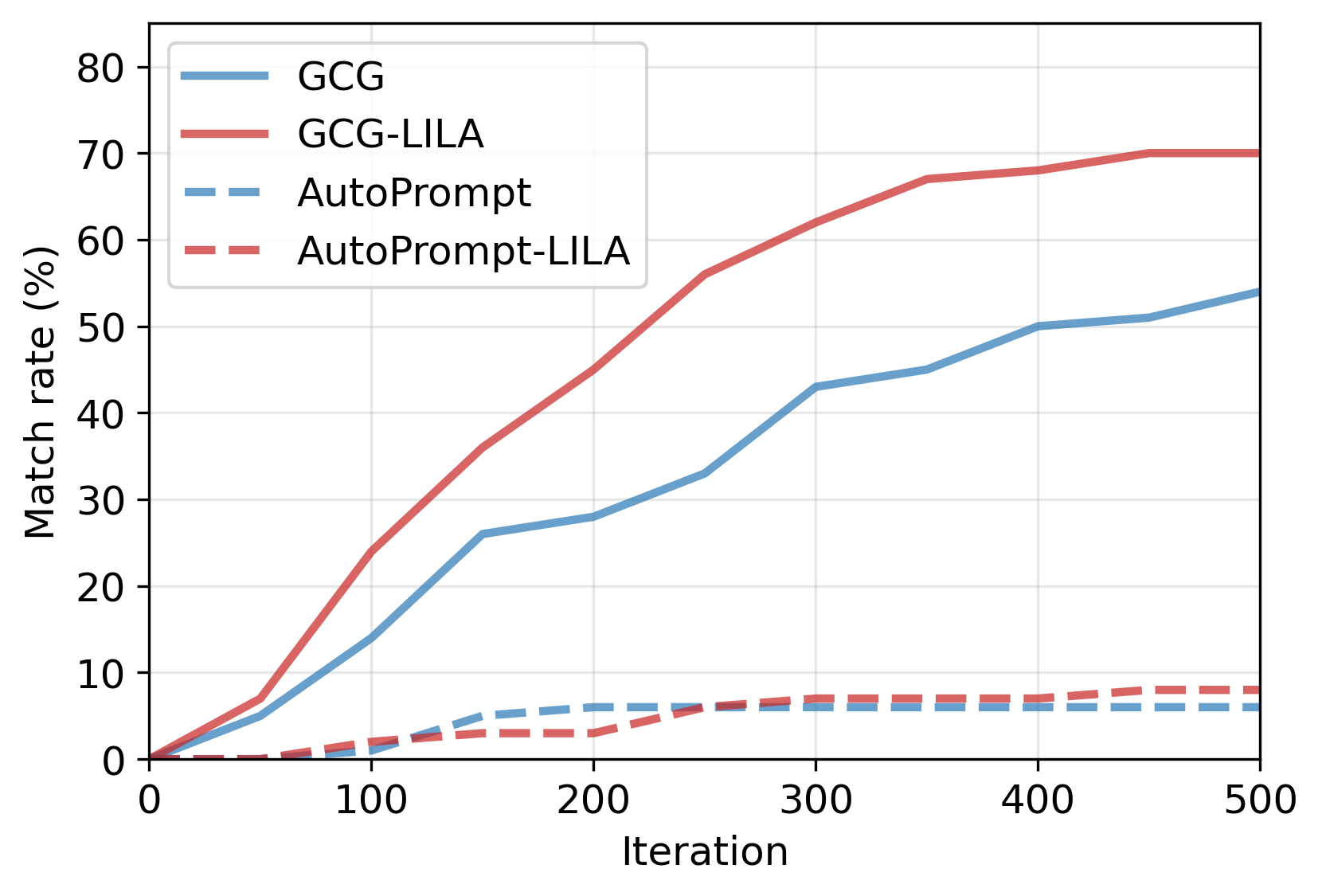}} 
    \vskip0.04in
    \caption{How (a) the loss and (b) the match rate changes with attack iterations. The attacks are performed against Llama-2-7B-Chat model to generate query-specific adversarial suffixes on AdvBench. Best viewed in color.}
    \label{fig:ila}
    \end{minipage}
    \vskip-0.25in
\end{figure}

Nevertheless, by unfolding the internal computation of an LLM as a grid of token representations $\{h_{r,o}\}$, where $h_{r,o}$ denotes the $o$-th token representation at $r$-th layer, as depicted in Figure~\ref{fig:grid}, some previous work~\cite{meng2022locating, meng2022mass, ghandeharioun2024patchscope, nostalgebraist2020logitlens} have shown that different token representations at a intermediate layer show distinct effects on the model output.
It inspires us to further investigate whether the positive correlation exists between the scalar projection of a single token representation onto its directional guide and the adversarial loss (\ie, $\mathrm{PCC} (h^T_{r,o} v_{r,o}, L(x))$), and whether this correlation is stronger than the positive correlation between the scalar projection computed on its gradient and the adversarial loss (\ie, $\mathrm{PCC} (h^T_{r,o} \nabla_{h_{r,o}}L(x), L(x))$).
In Figure~\ref{fig:5.1}, we present $\mathrm{PCC} (h^T_{r,o} v_{r,o}, L(x))$ and $\mathrm{PCC} (h^T_{r,o} \nabla_{h_{r,o}}L(x), L(x))$ with different choices of $o$-th token at the mid-layer of Llama-2-7B-Chat.
Since the number of target tokens varies across different adversarial prompts, we select the first five tokens from the target phrase.
First, the scalar projections computed using the directional guide and intermediate gradient both show a very weak positive correlation with the adversarial loss in the intermediate representations of adversarial tokens.
Second, the PCCs calculated for the token representation of ``]'', which is the last token of the input prompt and is expected to output the first token of the target phrase (\eg, ``Sure''), demonstrate that the scalar projection computed using the gradient can better indicate the change in adversarial loss than that computed using the directional guide.
We consider that successfully generating the first token of the target (e.g., ``Sure'') plays a crucial role in reducing the adversarial loss. The adversarial prompt used to obtain the directional guide is not sufficiently optimized to minimize the loss associated with generating the first token of the target.
Somewhat interestingly, the high PCC value between the projection of last token representation onto the gradient (or directional guide) and the adversarial loss indicates that directly moving the last token representation along the direction of gradient (\ie, intervening in the last token representation by adding a vector in the direction of gradient or directional guide) will effectively reduce the adversarial loss.
Some concurrent work~\cite{arditi2024refusal, xu2024uncovering} also observes that certain directions can be added to the token representations to jailbreak safety-aligned LLMs. In contrast to our approach, they collect the token representations of a set of harmful queries and a set of harmless queries, defining the direction as the difference between the average last token representations of these two sets.
Third, for other token representations, the scalar projections onto directional guides show a stronger correlation with adversarial loss than the scalar projections onto their gradients. This indicates that projecting the token representation onto the directional guide can better reflect changes in adversarial loss compared to projecting it onto the intermediate gradient.

Therefore, to effectively adapt ILA for adversarial prompt generation we propose replacing the intermediate gradient of the adversarial loss with a directional guide at specific token representations, rather than at the entire intermediate representation as in the original implementation. 
We re-normalize the directional guide by $\|\nabla_{h_{r,o}} L(x)\|_2 / \|v_{r,o}\|_2$, in order to avoid too large (or small) compared to the gradient \wrt other token representations. 
We denote this adaptation of ILA as Language ILA (LILA).
When performing LILA at the $o$-th token representation at $r$-th layer, during backpropagation, the gradient \wrt the token representation, \ie, $\nabla_{h_{r}} L(x)$ is changed to 
\begin{equation}
    \nabla^{\mathrm{LILA}}_{h_{r}} L(x) = [\nabla_{h_{r,1}} L(x), \nabla_{h_{r,2}} L(x), ..., \frac{\|\nabla_{h_{r,o}} L(x)\|_2}{\|v_{r,o}\|_2} v_{r,o}, ..., \nabla_{h_{r,n^*-1}} L(x)].
\end{equation}
Another issue of the original ILA here is that it requires a preliminary attack, which is time-consuming to obtain in the setting of attacking LLMs.
We make a compromise by utilizing the current adversarial prompt to obtain the directional guide, which leads to little increase in computational complexity. 
For instance, at $t$-th iteration, the directional guide is $v^t_{r,o} = h^0_{r,o} - h^t_{r,o}$.
Note that we only modify the gradient computation step in each iteration. Therefore, the step of evaluating candidate adversarial suffixes can help reduce the adversarial loss, providing useful directional guidance in the initial iterations.
We evaluate LILA using GCG and AutoPrompt as baseline attacks against the Llama-2-7B-Chat model on AdvBench. As shown in Figure~\ref{fig:ila}, LILA demonstrates a lower adversarial loss and a higher match rate compared to the baseline attacks. 
The experimental results suggest that our adaptation of ILA indeed improves discrete optimization in gradient-based adversarial prompt generation and offers insights to more effectively address similar discrete optimization problems within LLMs.

\section{Experiments}
We introduce the evaluation metrics in Section~\ref{sec:metrics}, and then present the experimental results in Section\ref{sec:results}. 
Some detailed experimental settings and ablation studies are presented in the Appendix.

\subsection{Metrics}
\label{sec:metrics}
We use two metrics for the evaluations: match rate (MR) and attack success rate (ASR).
The match rate counts the fraction of adversarial examples that make the output exactly match the target string, and was used to evaluate different optimization methods in the paper of GCG~\cite{zou2023universal}.
The attack success rate is evaluated using the evaluator proposed by HarmBench~\cite{mazeika2024harmbench}, which achieves over $93\%$ human agreement rate as reported in their paper.
We set the models to generate 512 tokens with greedy decoding during the evaluation phase.
Note that for the universal adversarial suffix generation, we observed that the ASRs of the adversarial suffixes obtained by multiple runs for the same method differ significantly. 
Hence, we run each method ten times and report not only the average ASR (AASR) but also the best ASR (BASR) and the worst ASR (WASR).

\subsection{Experimental Results}
\label{sec:results}

Following the evaluations by GCG~\cite{zou2023universal}, we perform evaluations in the settings of query-specific adversarial suffix generation and universal adversarial suffix generation.
For query-specific adversarial suffix generation, following~\cite{zou2023universal}, we use first 100 harmful behaviors in AdvBench.
For universal adversarial suffix generation, we use the first 10 harmful queries in AdvBench to generate a universal adversarial suffix and test it on the rest 510 harmful queries in AdvBench.
We also test our method for universal adversarial suffix generation on the standard behaviors set from HarmBench~\cite{mazeika2024harmbench}, following the setting suggested in HarmBench.
We use a Top-$k$ selection of 4 and a candidate set size of 20 for Llama and Mistral models, and a Top-$k$ selection of 64 and a candidate set size of 160 for Phi3-Mini-4K-Instruct.
Since Llama-2-Chat~\cite{touvron2023llama} models show great robust performance to gradient-based adversarial prompt generation~\cite{mazeika2024harmbench}, we mainly evaluate the methods on the model of Llama-2-7B-Chat~\cite{touvron2023llama} and the model of Llama-2-13B-Chat~\cite{touvron2023llama}. In addition, Mistral-7B-Instruct-v0.2~\cite{jiang2023mistral} and Phi3-Mini-4K-Instruct~\cite{abdin2024phi} are also considered.

The comparison results in the setting of query-specific adversarial suffix generation are shown in Table~\ref{tab:individual}. 
Experimental results demonstrate that both LSGM and LILA outperform the GCG attack on three safety-aligned LLMs. 
Our combination method further boosts both the match rates and the attack success rates, achieving the best performance. Specifically, GCG-LSGM-LILA achieves gains of $+30\%$, $+19\%$, $+19\%$, and $+21\%$ when attacking the Llama-2-7B-Chat and Llama-2-13B-Chat, Mistral-7B-Instruct, and Phi3-Mini-4K-Instruct, respectively.
Moreover, since these methods reduce the gap between the input gradients and the effects of loss change results from token replacements, the size of candidate set at each iteration can be reduced for saving running time.
We show the results of GCG with the default setting described in their paper, which evaluates 512 candidate adversarial suffixes at each iteration.
It can be seen that our combination still shows outstanding performance, by only using $4\%$ time cost compared with the GCG with their initial setting (denoted as GCG$^*$ in the table).

\begin{table*}[h]
\caption{
Match rates, attack success rates, and time costs for generating query-specific adversarial suffixes on AdvBench are shown. The symbol $^*$ indicates the use of the default setting for GCG, \ie, using a Top-$k$ of 256 and a candidate set size of 512. Time cost is derived by generating a single adversarial suffix on a single NVIDIA V100 32GB GPU.
}
\vskip-0.1in
\label{tab:individual}
\begin{center}
\resizebox{0.97\linewidth}{!}{
\renewcommand{\arraystretch}{1.1}
\begin{tabular}{lcccccccccccc}
\toprule
\multirow{2}{*}{Method}    & \multicolumn{3}{c}{Llama-2-7B-Chat} & \multicolumn{3}{c}{Llama-2-13B-Chat} & \multicolumn{3}{c}{Mistral-7B-Instruct} & \multicolumn{3}{c}{Phi3-Mini-4K-Instruct}  \\\cmidrule(lr){2-4} \cmidrule(lr){5-7} \cmidrule(lr){8-10} \cmidrule(lr){11-13}
& MR            & ASR             & Time  & MR                    & ASR          & Time & MR                       & ASR               &  Time              & MR                       & ASR               &  Time        \\\midrule
GCG*           & 60\%          & 44\%            &  85m          & 58\%                 & 40\%        &  170m  & \textbf{95\%}            & 92\%              & 85m        & 70\%                       & 61\%               &  50m              \\\midrule
GCG            & 54\%          & 38\%            &  3m          & 37\%                 & 33\%         &   6m   & 73\%                     & 74\%              & 3m         & 60\%                       & 59\%               &  17m                \\
GCG-LSGM       & 72\%          & 62\%            &  3m          & 52\%                 & 43\%         &    6m    & 93\%                     & 88\%              & 3m       & 75\%                       & 64\%               &  17m                  \\
GCG-LILA       & 70\%          & 59\%            &  3m          & 52\%                 & 48\%         &    6m    & 83\%                     & 80\%              & 3m       & 65\%                       & 61\%               &  17m                  \\
GCG-LSGM-LILA  & \textbf{87\%}  & \textbf{68\%}  &  3m          & \textbf{62\%}        & \textbf{52\%} &   6m    & 94\%                    & \textbf{93\%}     & 3m        & \textbf{81\%}              & \textbf{68\%}      &  17m                  \\

\bottomrule
\end{tabular}} 
\end{center} \vskip -0.1in
\end{table*}

The results of the attacks in the setting of universal adversarial suffix generation are shown in Table~\ref{tab:multiple}.
It can be observed that our combination not only achieves a remarkable improvement in the average ASR but also enhances both the worst and best ASRs obtained over 10 runs. Specifically, when attacking Llama-2-7B-Chat model on AdvBench, the GCG-LSGM-LILA achieves an average ASR of $60.32\%$, which gains a $+33.64\%$ improvement compared with the GCG attack. Moreover, for the worst and best ASR, GCG-LSGM-LILA achieves gains of $+34.82\%$ and $+31.06\%$, respectively.

\begin{table*}[h]
\caption{
Attack success rates for generating universal adversarial suffixes on AdvBench and HarmBench. 
The average ASR (AASR), the worst ASR (WASR), and the best ASR (BASR) are obtained by performing each attack ten times.
}
\vskip-0.1in
\label{tab:multiple}
\begin{center}
\resizebox{0.85\linewidth}{!}{
\renewcommand{\arraystretch}{1.1}
\begin{tabular}{ccccccccc}
\toprule
\multirow{2}{*}{Dataset}  & \multirow{2}{*}{Model} & \multicolumn{3}{c}{GCG}     & \multicolumn{3}{c}{GCG-LSGM-ILA} \\\cmidrule(lr){3-5}\cmidrule(lr){6-8}
                          &                           & AASR    & WASR    & BASR    & AASR         & WASR         & BASR         \\\midrule
\multirow{4}{*}{AdvBench} &    Llama-2-7B-Chat        & 26.68\% & ~0.40\%  & 55.80\% & \textbf{60.32\%}      & \textbf{35.22\%}      & \textbf{86.86\%}      \\
                          &    Llama-2-13B-Chat       & 20.98\% & ~0.00\%  & 37.06\% & \textbf{45.27\%}      & \textbf{~~7.45\%}      & \textbf{67.25\%}      \\
                          &    Mistral-7B-Instruct    & 56.53\% & 34.51\% & 92.16\% & \textbf{73.48\%}      & \textbf{50.80\%}        & \textbf{92.25\%}      \\
                          &    Phi3-Mini-4K-Instruct  & 35.84\% & 20.39\% & 46.27\% & \textbf{44.80\%}      & \textbf{31.18\%}        & \textbf{61.18\%}      \\\midrule
\multirow{4}{*}{HarmBench} &    Llama-2-7B-Chat        & 56.90\% & 33.00\% & 66.50\% & \textbf{69.35\%}      & \textbf{57.50\%}      & \textbf{87.00\%}      \\
                          &    Llama-2-13B-Chat       & 37.40\% & 13.50\%  & 64.50\% & \textbf{53.55\%}      & \textbf{22.00\%}      & \textbf{81.50\%}      \\
                          &    Mistral-7B-Instruct    & 75.00\% & 37.50\% & 90.50\% & \textbf{81.00\%}      & \textbf{66.50\%}        & \textbf{93.00\%}      \\
                          &    Phi3-Mini-4K-Instruct  & 49.90\% & 29.00\% & 65.50\% & \textbf{63.80\%}      & \textbf{48.50\%}        & \textbf{80.50\%}      \\
\bottomrule

\end{tabular}} 
\end{center} 
\vskip -0.1in
\end{table*}

We also test the transfer attack performance of our method. We use the universal suffixes generated by performing GCG and GCG-LSGM-LILA against Llama-2-7B-Chat to attack GPT-3.5-Turbo on 100 harmful queries in AdvBench. The test queries are distinct from the queries that are used to generate universal suffixes. The results are shown in Table~\ref{tab:transfer}. It can be observed that our GCG-LSGM-LILA achieves remarkable improvements in the average, worst, and best ASRs obtained over 10 runs.

\begin{table*}[h]
\caption{
The performance of transfer attack against GPT-3.5-Turbo on AdvBench. 
The average ASR (AASR), the worst ASR (WASR), and the best ASR (BASR) are obtained by performing each attack ten times.
}
\vskip-0.1in
\label{tab:transfer}
\begin{center}
\resizebox{0.45\linewidth}{!}{
\renewcommand{\arraystretch}{1.1}
\begin{tabular}{lccc}
\toprule
Method        & AASR    & WASR & BASR \\\midrule
GCG           & 38.30\% & 24\% & 48\% \\
GCG-LSGM-LILA & \textbf{45.20\%} & \textbf{35\%} & \textbf{81\%} \\
\bottomrule
\end{tabular}} 
\end{center} 
\vskip -0.1in
\end{table*}

\section{Conclusions}

  In this paper, we present a new perspective on the discrete optimization problem in gradient-based adversarial prompt generation. That is, using gradient \wrt the input to reflect the change in loss that results from token replacement resembles using input gradient calculated on the substitute model to indicate the real effect of perturbing inputs on the prediction of a black-box victim model, which has been studied in transfer-based attacks against image classification models. By making some appropriate adaptations, we have appropriated the ideologies of two transfer-based methods, namely, SGM and ILA, into the gradient-based white-box attack against LLMs. Our analysis of the mechanisms behind their effective performance has provided new insights into solving discrete optimization problem within LLMs. Furthermore, the combination of these methods can further enhance the discrete optimization in gradient-based adversarial prompt generation. Experimental results demonstrate that our methods significantly improves the attack success rate for both white-box and transfer attacks.

{\small
\bibliographystyle{plain}
\bibliography{ref}
}








\newpage
\appendix


\section{Experimental Settings}
\label{sec:setting}



\textbf{Hyper-parameters.} 
For SGM, we simply set $\gamma=0.5$.
For the selection of intermediate representation to perform LILA, we choose the first token from the target phrase since it empirically performs better than selecting other tokens or all tokens except the adversarial suffix and the last input token.
For the intermediate layer, we choose the midpoint of the model layers.
For instance, for Llama-2-13B-Chat~\cite{touvron2023llama}, we select the 20-th layer, whereas for other models with 32 layers, we select the 16-th layer. 
We perform 500 iterations for all methods, with the number of adversarial tokens set to 20.

\section{Ablation Study}
\label{sec:ablation}

\begin{figure}[h]
\centering 
 \subfigure[LSGM]{\label{fig:gamma}
 \includegraphics[width=0.4\textwidth]{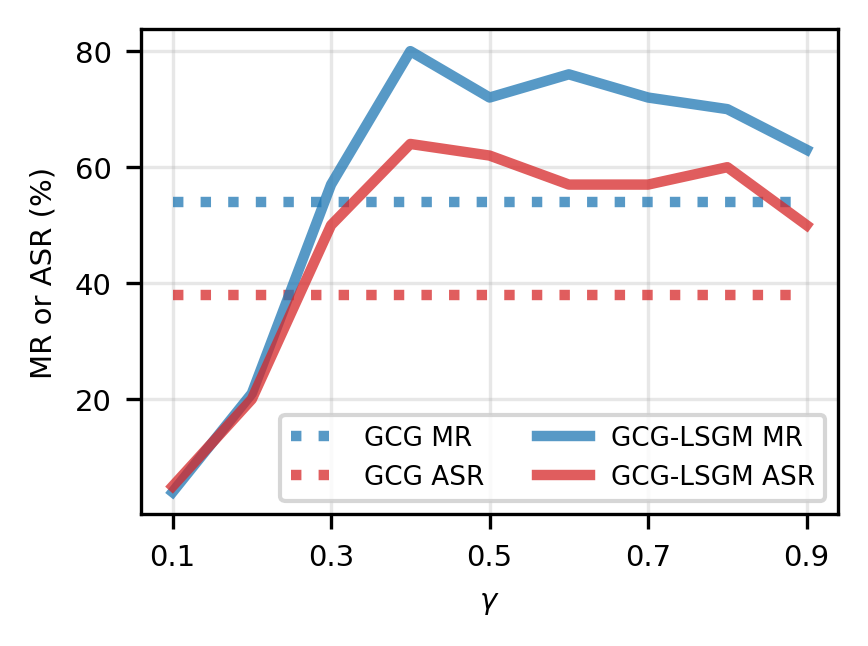}}
 \subfigure[LILA]{\label{fig:ila_ce_layer}
 \includegraphics[width=0.4\textwidth]{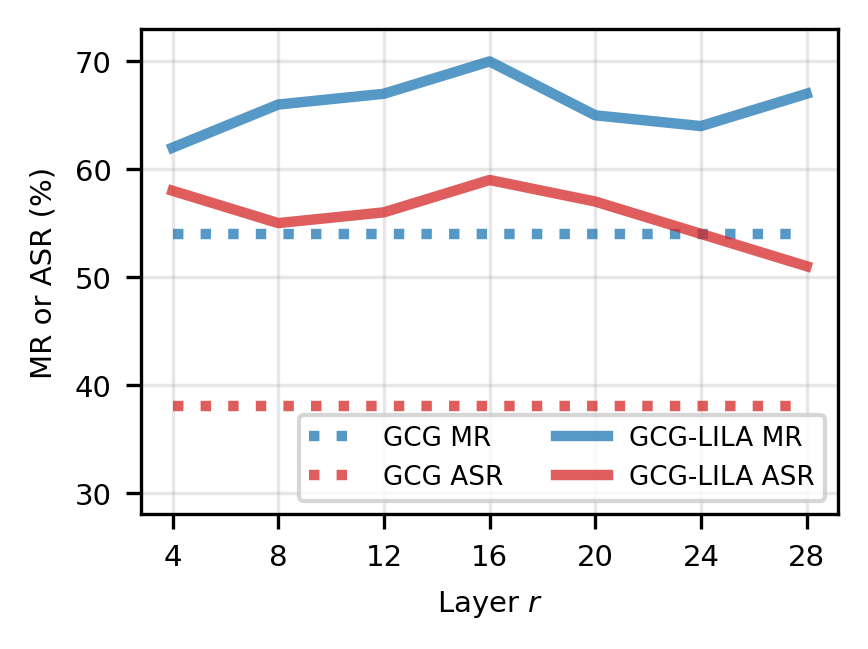}}
\caption{How the match rate and attack success rate change with (a) the choice of $\gamma$ for GCG-LSGM, (b) the choice of layer for GCG-LILA. Best viewed in color.}
\end{figure}

We evaluate GCG-LSGM with varying the choice of $\gamma$ in Figure~\ref{fig:gamma}, and GCG-LILA with varying the choices of the layer in Figure~\ref{fig:ila_ce_layer}, in the setting of query-specific adversarial suffix generation for attacking Llama-2-7B-Chat on AdvBench. 
It can be seen that for LSGM, a large range of the choice of $\gamma$ ($0.3 \sim 0.9$) leads to improved performance.
For LILA, it demonstrates consistently more effective performance on all choices of the layer.

\section{Limitations}
We use a model-based evaluator provided by a benchmark, namely, HarmBench~\cite{mazeika2024harmbench}, to evaluate the success rates of attacks. Although it achieves a high human agreement rate ($>93\%$), it cannot perfectly judge whether an attack is successful. More accurate evaluators will be adopted in future work to better evaluate attack performance.

\section{Broader Impacts}
This work has improved the effectiveness of automatic adversarial prompt generation on LLMs. Specifically, on attacking Llama-2-Chat models, which are considered to be highly robust, our work achieves over $+30\%$ gains compared with GCG attack. This may aid in assessing of the robustness of safety-aligned LLMs, and could potentially be used in adversarial training method to improve the robustness of LLMs. Moreover, our research may also contribute to solving discrete optimization problem within LLMs, \eg, prompt tuning.

\end{document}